\documentclass[doublecol]{epl2} 
\usepackage{amsmath}
\usepackage{amssymb}
\title{Synchronization on directed small worlds: feed forward loops and cycles.}
\shorttitle{Synchronization on directed Small Worlds: The role of motifs} 

\author{Markus Brede\inst{1} }
\shortauthor{M. Brede}

\institute{                    
  \inst{1} CSIRO Marine and Atmospheric Research, CSIRO Centre for Complex System Science, F C Pye Laboratory - GPO Box 1666, Clunies Ross Street
Canberra ACT 2601, Australia\\

}

\abstract{In this paper we investigate the influence of directed motifs on the synchronization properties of Kuramoto oscillators on directed networks. Building different types of sparse directed small world networks similar to the Watts and Strogatz procedure we establish that feed forward loops favour synchronization on directed networks. The paper highlights the importance of local network characteristics for synchronization.}

\begin{document}

\maketitle

\section{Introduction}

Starting with Watts and Strogatz introduction of the small world paradigm \cite{SW} synchronization phenomena on complex networks have found much attention in the recent literature. A question of pervading interest in the field is how features of the network architecture relate to characteristics of the dynamics on the network, e.g. the onset of synchronization or the achievement of full synchronization. So far much effort has been devoted to an exploration of the stability of the fully synchronized state via the master stability function (MSF) approach \cite{Pecora1} which has allowed considerable insight about general structural requirements of network topologies to synchronize (see, e.g., \cite{Nishi,Hwang,Motter,Motter1,Chavez,Donetti,MB0}). One important finding is that the variance of the in-degree distribution of the network is a strong determinant of the synchronizablity \cite{Donetti,Motter,Motter1}, the more in-degree homogeneous networks being easier to synchronize than more heterogeneous ones. Other studies have demonstrated that small distances, low clustering, disassortative degree mixing and an even distribution of loads on nodes tend to enhance the stability of the fully synchronized state. However, while allowing general insights about a broad class of systems, the MSF approach is limited to systems of symmetrically coupled \cite{Remark0} identical oscillators and properties of the fully synchronized state.

Studying the transition to synchronization in the Kuramoto system \cite{Kuramoto} in more detail, a number of recent studies has shown that the synchronization properties can depend on the whole range of coupling strengths \cite{Gardenes,Gardenes1}. Networks may exhibit an onset of synchronization for very small coupling, but the transition to full synchronization may only occur for very large coupling. Insights in these studies have generally been obtained by a finite size scaling analysis of the synchronization transition \cite{Hong,Gardenes1,Gardenes2}. Other recent studies have explored the role of the oscillator placement for synchronization \cite{MB1,MB2,MB3}.

Notably, most studies of synchronization on networks have focussed on undirected networks. In many cases, however, real-world networks are inherently directed. Thus, they are characterized by a number of features, as, e.g., substantially different in- and out-degree sequences, a variety of in- and out-degree correlations, or a more subtle component organization, that are not found in undirected networks. One might expect that these characteristics will contribute to the richness of dynamical phenoma on such networks. Indeed, this has been confirmed in a couple of recent studies \cite{Timme,MB1}  that started to explore the relationship between structural features of directed networks and synchronization properties. One finding of this work is that the component structure of directed networks strongly influences synchronization properties \cite{Timme}. Further, in \cite{MB1} a connection between various degree correlations and synchronization has been made.

Another observation that adds to the importance of studying synchronization on directed networks is that the synchronized state can generally be achieved for less coupling, i.e. less 'cost'. This may be relevant for technological applications.

In this paper, we build on these insights and explore the influence of small subgraphs or motifs (for a definition of network motifs see \cite{Milo}) for a network's synchronization properties. Specifically, we study the transition towards synchronization in the Kuramoto model on directed small world networks that have been constructed in such a way that different (directed) motifs are over- or underrepresented, while other relevant network characteristics have been kept fixed. This allows to make a strong point that directed motifs, i.e. features of the local organization of directed networks, can strongly influence the transition towards synchronization.

\section{Model Description}

More specifically, we study the Kuramoto system
\begin{equation}
\label{E1}
 \dot{\phi}_i=\omega_i+\sigma \sum_j a_{j i} \sin(\phi_j-\phi_i),
\end{equation}
where the $\phi_i,i=0,...,N-1$ are the phases and the $\omega_i$ the native frequencies of the limit-cycle oscillators, $\sigma$ is the coupling strength, and the binary coupling matrix $a_{ij}\in \{0,1 \}$ represents the interaction network. For all the following experiments, native frequencies are always drawn uniformly at random from the interval $[-1,1]$. Because of its simplicity the Kuramoto model is a suitable choice for the dynamics.

\begin{figure}[tbp]
 \includegraphics[width=.45\textwidth]{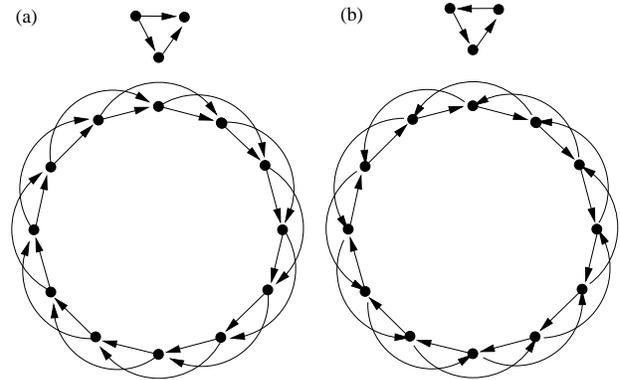}%
 \caption{Illustration of the directed substrate ring graphs to construct directed small worlds. (a) A ring graph of range 2 built from feed forward loops and (b) a ring graph of range 2 built from cycles. Note, that both graphs are strongly connected and have the same in- and out-degree sequences. Small worlds built from cycles are smaller than small worlds build from the feed forward loop.}
\label{fig.1}
 \end{figure}

Following a very similar procedure to that of Ref. \cite{SW}, the interaction networks are constructed to be small worlds. For this, we start with two different directed ring graph substrates, depicted in Fig. \ref{fig.1}. The ring graph model (a) is built from triangles that are feed forward loops, while the model (b) builds a ring graph from cycles. More specifically, to construct ring graphs of range $r$ we set
\begin{equation}
\label{R1}
 a_{ji}=\sum_{k=1}^r \delta_{j, i-k}
\end{equation}
for model (a) and
\begin{equation}
 \label{R2}
 a_{ji}=\delta_{j,i-1}+ \sum_{k=1}^{r-1} \delta_{j,i+k+1} 
\end{equation}
for model (b) and $i,j=1,...,N=2n$. In Eq.'s (\ref{R1}) and (\ref{R2}) $\delta_{i,j}=1$ if $i=j$ and $\delta_{i,j}=0$ if $i\not=j$ denotes the usual Kronecker delta and all indices are understood modulo $N$.

It is important to note that both graphs are strongly connected, have the same in- and out-degree sequences and the same correlations between degrees as well as the same triangle densities, i.e. the same (undirected) clustering coefficients.
Note, however, that the average distances between nodes are different on both substrate graphs. Essentially, cycles allow a walker to also step `backwards', leading to substantially smaller average and maximum distances $d_\text{max}^{(b)}\sim N/3$ between nodes for the model (b) in comparison to the model (a) for which $d_\text{max}^{(a)}\sim N/2$.

The construction idea of the ring graphs from loops of length three can easily be generalized to loops of arbitrary length. For looplength $l$ one may define
\begin{align}
 \label{L1}
 a_{ji}=\delta_{j,i-1}+\sum_{k=1}^{r-1} \delta_{j,i-k(l-1)}
\end{align}
for a ring graph built from feed forward loops of length $l$, i.e. model (a), and
\begin{align}
 \label{L2}
a_{ji}=\delta_{j,i-1}+\sum_{k=1}^{r-1} \delta_{j,i+k(l-1)},
\end{align}
$i,j=1,...,N=(l-1)(r-1)n$, for a ring graph built from cycles of length $l$, see model (b). This construction results in substrate graphs with loops no shorter than length $l$. One notes that in this more general situation distances decrease with increasing loop length $l$, but are still larger for model (a) where $d_\text{max}^{(a)}\sim N/((r-1)(l-1))$ in comparison to those for model (b), where $d_\text{max}^{(b)}\sim N/((r-1) (l-1)+1)$. Below we restrict ourselves to graphs with range $r=2$. We have, however, also explored other values of the range only to find no qualitiative difference to the general results presented.

Another important point to note is that following \cite{Moreno} one can characterize the synchronizablity of local motifs by the coupling strength $\sigma^*$ for which the average probability that the oscillators on the nodes are synchronized exceeds $1/2$. Numerical simulations show that compared to feed forward loops cycles are the more synchronizable motif in this regard. Hence, based on shorter distances and higher synchronizability of the local motifs one should expect small worlds built from the configuration b) to be more synchronizable than those built from configuration a). As we will see below, this is however not the case.

From the substrate graphs small worlds are obtained in the following way. Iteratively, a node and one of its incoming links are picked randomly. The link is then rewired to a new randomly selected origin node, provided the resulting graph is still strongly connected \cite{Remark}. If the rewiring disconnects the strong component, a different link is picked instead. The procedure is repeated $pL$ times, where $L$ denotes the number of links in the graph. For small $p$ there are roughly $pL$ shortcut links in the system. As in \cite{SW} the parameter $p$ is the shortcut density. Increasing $p$ from $p=0$ gradually interpolates between a regular and a random graph topology. Note, however, that the graph is not fully random even for large $p$ since we have made sure of the strong connectedness and an equal number of incoming links for all nodes during the rewiring procedure \cite{remark2}. 

\section{Results}

The degree of synchronization between the oscillators of (\ref{E1}) is usually measured by the order parameter
\begin{equation}
\label{E2}
r(t)=\left|1/N \sum_j \exp{(i\phi_j(t))} \right|.
\end{equation}
If the phases of the $N$ oscillators are desynchronized one has $r\sim 1/\sqrt N$ while $r\sim {\cal O}(1)$ when a macroscopic part of the oscillators is synchronized. Increasing the coupling strength $\sigma$ between the oscillators one finds a second order phase transition. 

In our numerical experiments the order parameter (\ref{E2}) is determined by averaging over time (after allowing for a sufficient relaxtion time $T_\text{rel}=100...400$ depending on network size), and the randomness in the network, initial conditions $\phi_i(t=0)$ which are randomly drawn from $(-\pi,\pi]$, and the oscillator ensemble, i.e.
\begin{align}
\label{E4}
r= \langle 1/T_\text{rel}\sum_{t=T_\text{rel}}^{2T_\text{rel}} r(t)\rangle,
\end{align}
where brackets indicate the average over the different sources of randomness and $T_\text{rel}$ is the relaxation time.

\begin{figure*}[tbp]
 \includegraphics[width=.3\textwidth]{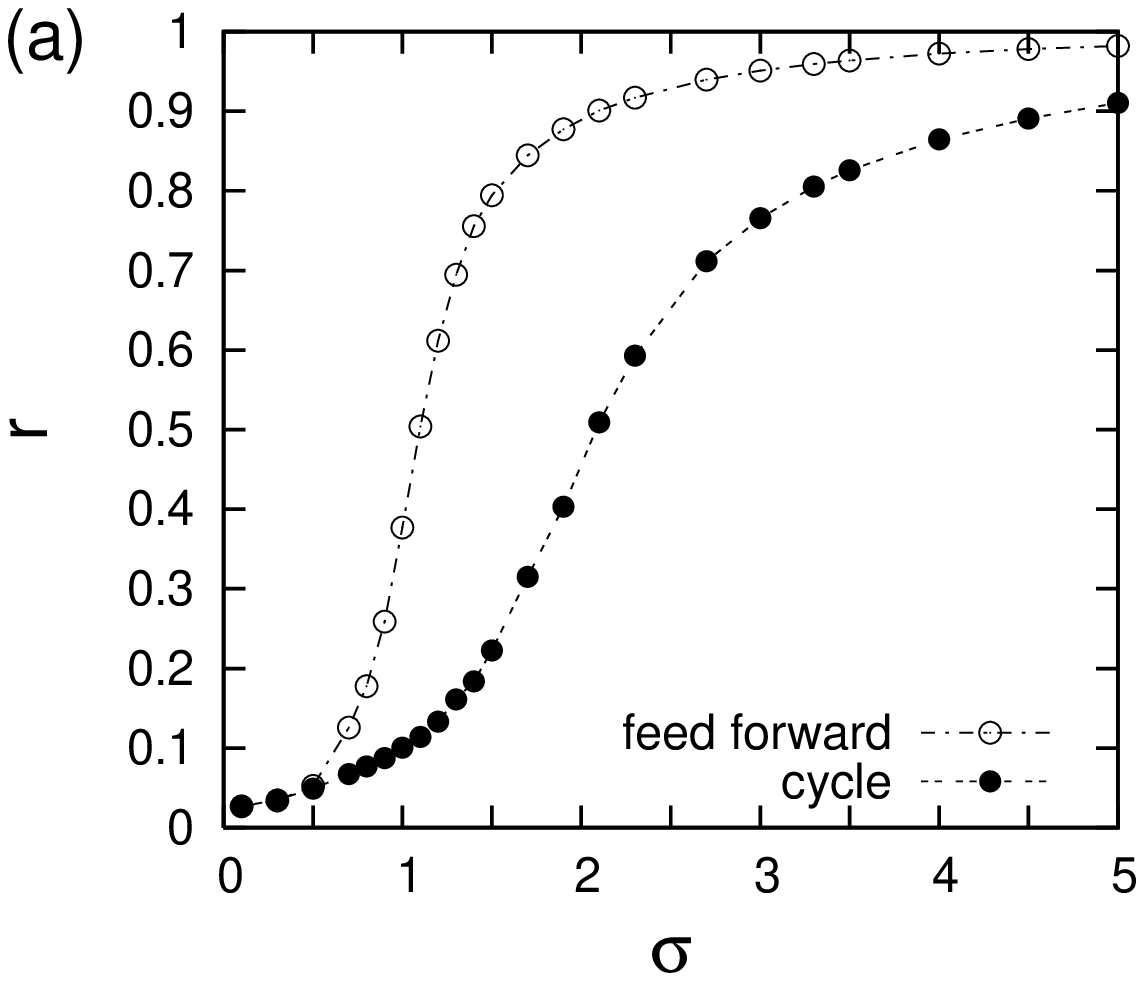}
 \includegraphics[width=.3\textwidth]{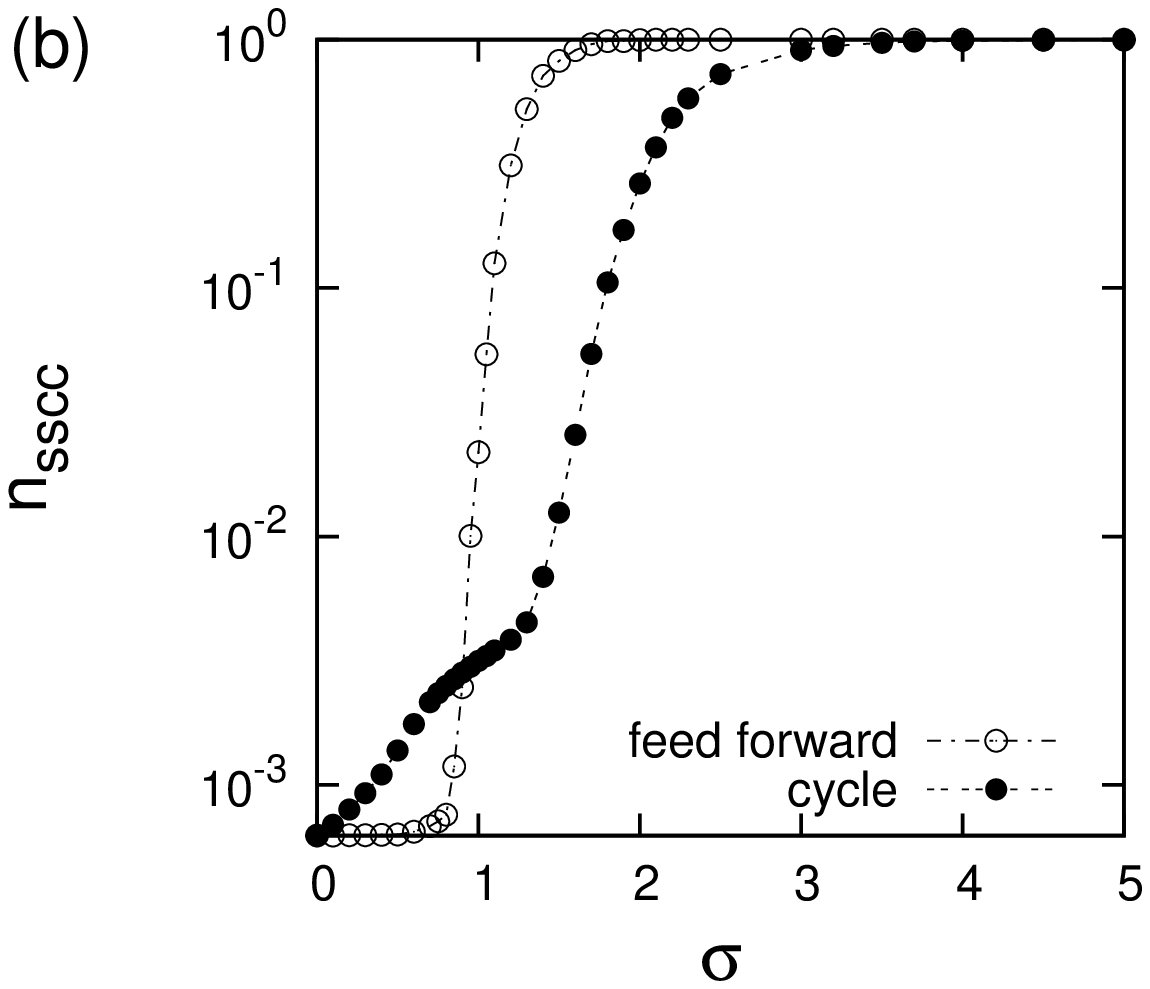}
 \includegraphics[width=.3\textwidth]{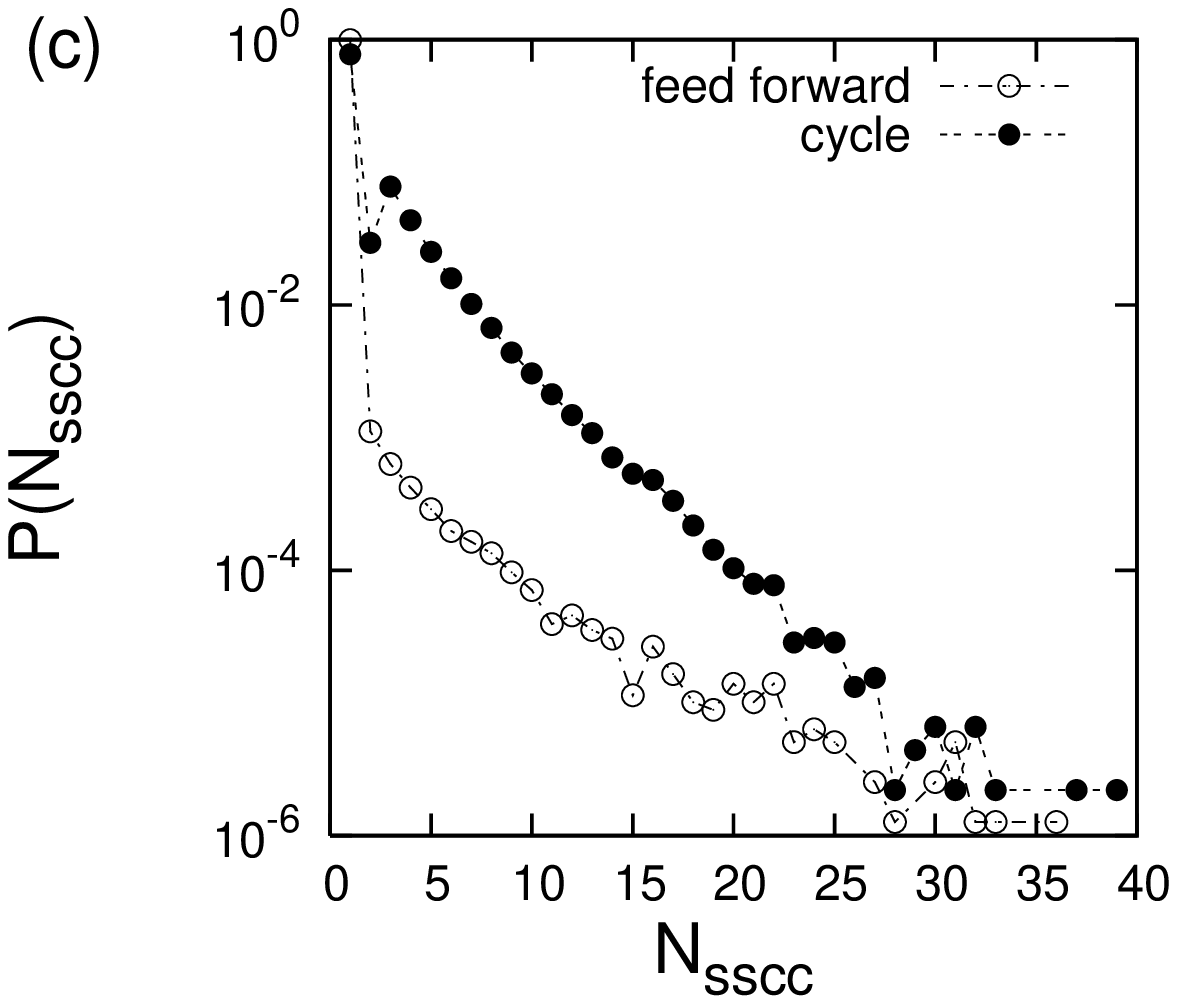}
 \caption{(a) Comparison of the transitions to synchronization on small worlds built from feed forward loops and from cycles. (b) Average relative size of the synchronized strongly connected components for both systems. (c) Distribution of the sizes of strongly connected synchronized components for $\sigma=.8$. Parameters are $N=1600$, $T_\text{rel}=300$, range $r=2$ and looplength $l=3$, density of shortcuts $p=0.05$.}
\label{fig.2}
\end{figure*}

Figure \ref{fig.2} illustrates some first numerical results for the synchronization transition on small world networks of size $N=1600$ with shortcut density $p=0.05$ built from the two different motifs. The data have been obtained for oscillators chosen uniformly at random from $[-1,1]$ as averages from a numerical integration of the system (\ref{E1}) over $200$ different randomly constructed small world networks built from the substrate graphs introduced above.
Clearly, the onset of synchronization as well the transition to full synchronization are achieved for lower coupling for the small worlds built from feed forward cycles.

To understand the difference in the synchronization behaviours between both systems, it is insightful to investigate the evolution of the sizes and the structures of synchronized components as the coupling strength $\sigma$ is increased. Synchronized components can be defined based on differences in the average frequencies $\overline{\omega}=\lim_{T\to\infty }1/T \int_0^T \dot{\phi}(t) dt$. Since the frequency resolution in our numerical simulations is given by $1/(2T_{\text{rel}})$, two oscillators 1 and 2 can be regarded as mutually entrained if $|\overline{\omega}_1-\overline{\omega}_2|<1/(2T_\text{rel})$. The synchronized in- or out-components of a given oscillator can then be defined as the entrained oscillators that can be reached following in- or out-links via entrained oscillators. Likewise, synchronized strongly connected components (SSCCs) are maximal sets of entrained oscillators that can reach each other by connected paths via entrained oscillators. In the following we denote the average size of SSCCs by $N_\text{sscc}$ and the average relative size by $n_\text{sscc}=N_\text{sscc}/N$.

Panel (b) of Fig.~\ref{fig.2} gives the evolution of the relative average size of SSCCs when the coupling $\sigma$ is increased. Crucially, as one may already expect from the construction of the substrate graphs, one notes that $N_\text{sscc}$ is very small almost up to the transition point for small worlds constructed by feed forward loops whereas $N_\text{sscc}$ grows gradually with increasing coupling for small worlds constructed from cycles. Moreover, as the distributions of the sizes of SSCCs for small coupling displayed in Fig. \ref{fig.2}c exemplify, there are many trivial and very small SSCCs in the first case, whereas many more large SSCCs of roughly comparable sizes are found in the second case. Global synchronization in the first case can thus be attained by a large SSCCs recruiting small or trivial SSCCs, whereas in the second case it arises from the competition and merger of SSCCs of comparable sizes. Moreover, as we observed before, the feed forward motif {\it per se} is less synchronizable than the cycle motif. Hence small synchronized components in the first case can be more easily broken up and recruited into larger synchronized clusters than in the second case. Superior synchronization on small worlds constructed by feed forward loops is thus attained, because the recruitment of small synchronized components into a large synchronized component can be achieved with less coupling than on small worlds built from cycles, where competition between large synchronized components of the same size exists.

Figure \ref{fig.2}, however, gives only a first indication of the synchronization properties and we support our first result by a more comprehensive set of experiments below. A more rigorous quantification of the synchronization transition can be obtained from a finite size scaling (FSS) analysis. In this, we follow the approach of \cite{Hong,Gardenes2} and assume the scaling ansatz
\begin{align}
\label{E3}
r(\sigma,N)=N^{-\alpha} F(N^{1/\nu} (\sigma-\sigma_c))
\end{align}
close to the critical point. In Eq. (\ref{E3}) $F(\cdot)$ is a universal scaling function and $\alpha$ and $\nu$ are critical exponents. The exponent $\nu$ describes the divergence of the correlation volume close to the critical point. From
\begin{align}
r(\sigma)\sim (\sigma-\sigma_c)^\beta
\end{align}
in the thermodynamic limit one obtains $\alpha=\beta/\nu$.
\begin{figure}[tbp]
 \includegraphics[width=.45\textwidth]{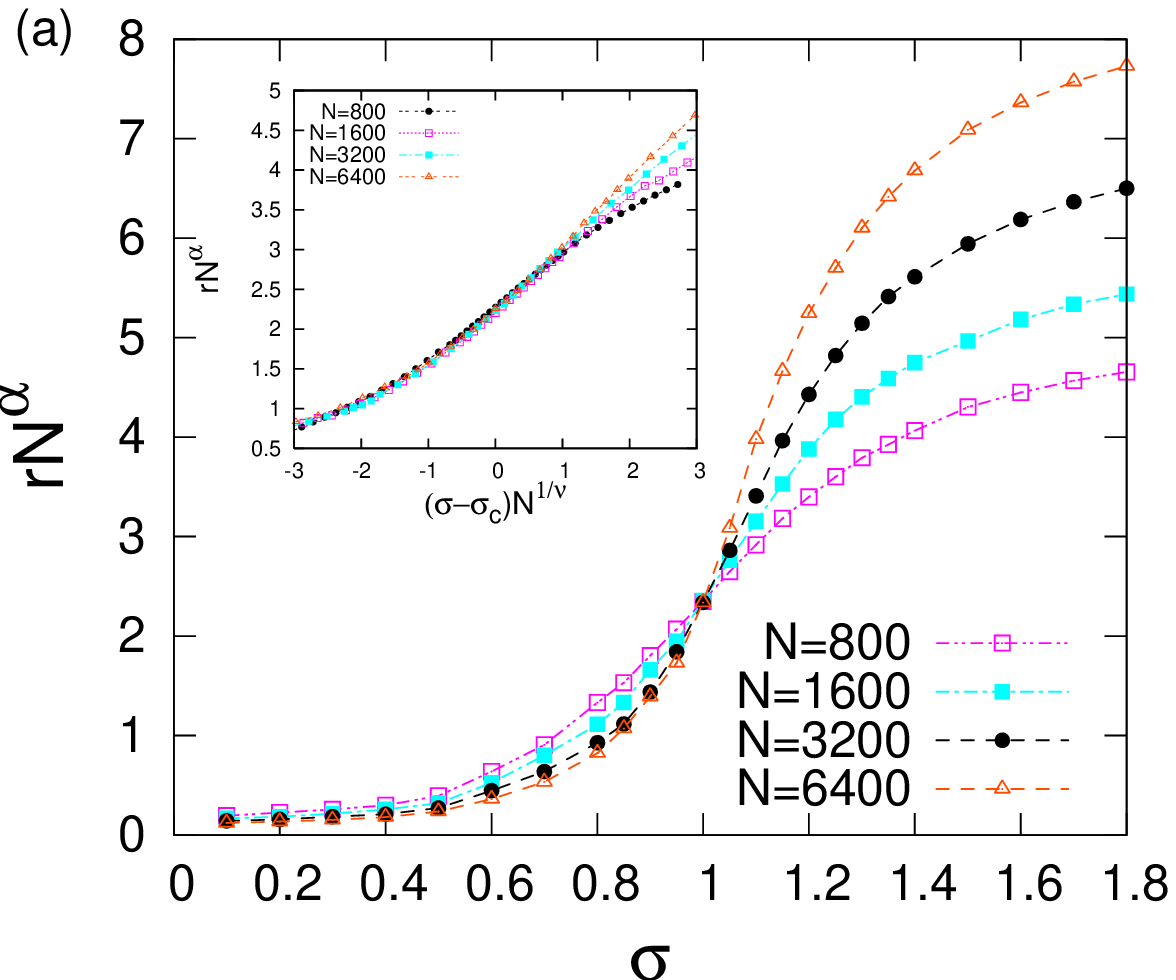}
 \includegraphics[width=.45\textwidth]{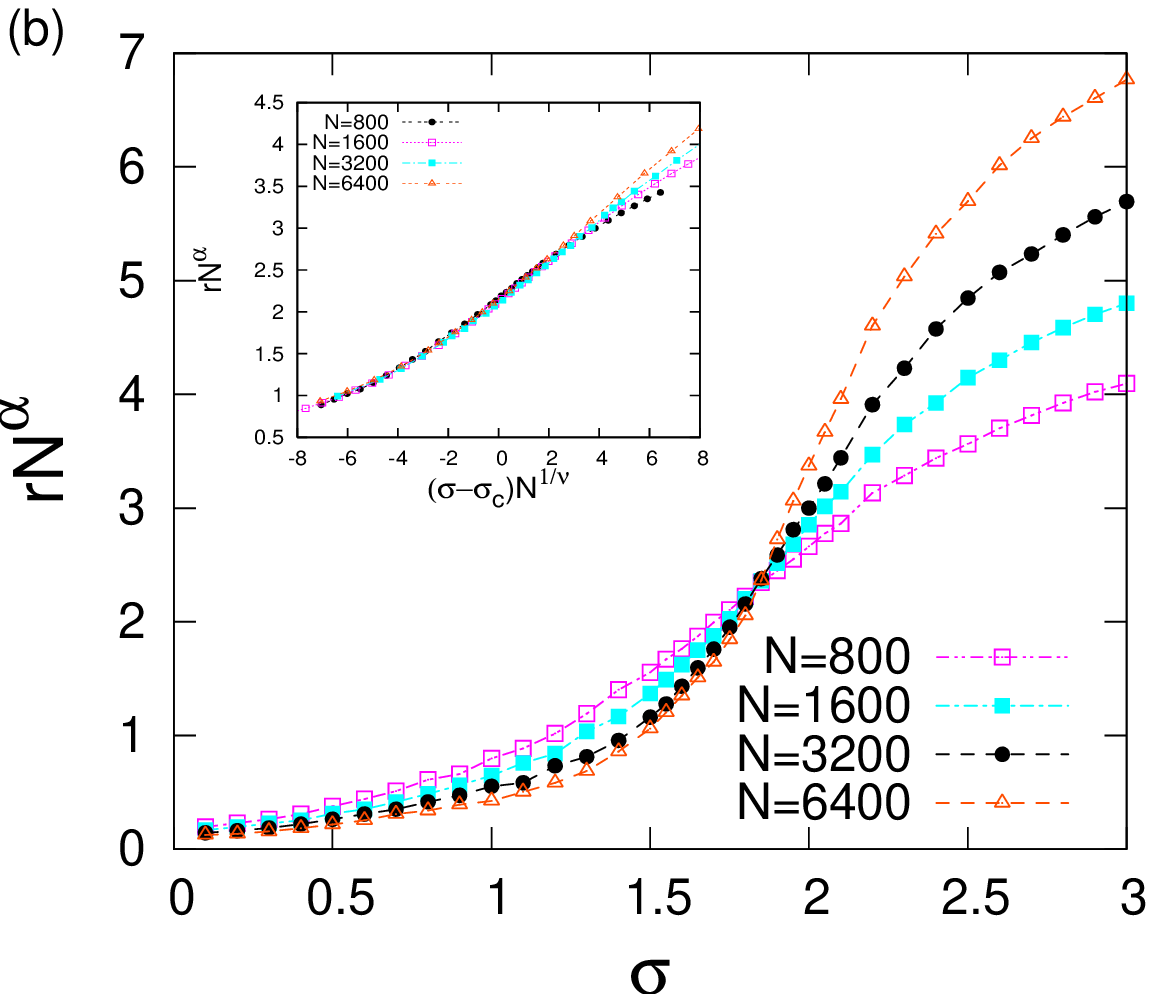}
 \caption{FSS analysis of the transition to synchronization for small world networks with short cut density $p=0.05$. (a) The transition for small worlds built from feed forward loops, $\alpha=.25$, $\sigma_\text{c}=.98(2)$. (b) The transition for small world networks built from cycles, $\alpha=.25$, $\sigma_\text{c}=1.78(2)$. The insets of the panels (a) and (b) show the collapse of data for different system sizes when plotting $Nr^\alpha$ vs. $(\sigma-\sigma_c)N^{1/\nu}$ for $\nu=3.0(1)$. Data points represent averages over at least 100 independent runs, error bars are smaller than the size of the data points. }
\label{fig.3}
\end{figure}
As in \cite{Hong,Gardenes2} the critical point can be obtained by plotting $rN^{\alpha}$ as a function of the control parameter $\sigma$ for different system sizes. From (\ref{E3}) one realizes that the curves for different $N$ intersect at the critical point for the right choice of the critical exponent $\alpha$. To determine the critical points we integrate Eq. (\ref{E1}) for several system sizes comprising up to $N=6400$ nodes. We then proceed with the FSS analysis. In Figure \ref{fig.3} an example of such an analysis is carried out for small worlds ($r=2,l=3$) with shortcut density $p=.05$ constructed from feed forward loops according to model (a) and cycles according to model (b), respectively. One finds that for both small world models for $\alpha=.25$ plotting $rN^\alpha$ vs. the coupling strength $\sigma$ data points for all system sizes intersect at a unique crossing point, the critical point $\sigma_\text{c}$. Having determined the critical coupling $\sigma_c$ and the exponent $\alpha$ one can then again use the FSS assumption of (\ref{E3}) to make all data points collapse to a single smooth curve (cf. insets of the panels of Fig. \ref{fig.3}). This allows the determination of the exponent $\nu$ as the exponent for which the best data collapse when plotting $rN^{\alpha}$ vs. $(\sigma-\sigma_c)N^{1\nu}$ is obtained. The analysis yields $\nu=3.0(1)$. One notes that the value of $\nu$ for both directed small world models is thus different from the value of $\nu$ determined for undirected small worlds in \cite{Hong}.

The more detailed analysis yields $\sigma_\text{c}=.98(2)$ for model (a) while $\sigma_\text{c}=1.78(2)$ for model (b). Thus, as one might already anticipate from Fig. \ref{fig.2} the synchronization transition on small world networks built from feed forward cycles occurs for substantially smaller coupling than for small world networks built from cycles. This appears particularly surprising, since small worlds built by cycles are typically much smaller than small worlds constructed from feed forward loops. In- and out degree sequences of the graphs and the triangle densities being the same, one concludes that the difference in the onset of synchronization is related to the local organization of the networks, i.e. the different motifs that they are constructed from.

\begin{figure}[tbp]
 \includegraphics[width=.45\textwidth]{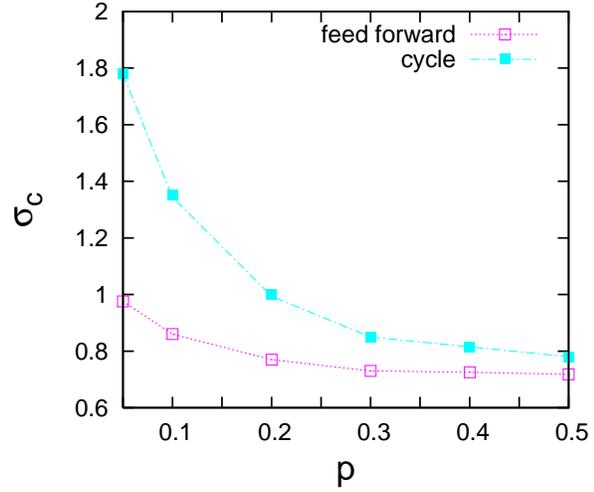} 
 \caption{Dependence of the critical point of the synchronization transition $\sigma_\text{c}$ on the shortcut density $p$ for small worlds with $r=2$ and $l=3$. The data are obtained from a FSS analysis, error bars are smaller than the symbols. Lines are just guides for the eye. }
\label{fig.4}
\end{figure}

This result proves robust if one varies the shortcut density of the small worlds, gradually tuning them towards random graphs, cf. Fig. \ref{fig.4}. For all shortcut densities we experimented with the FSS analysis yields $\sigma_\text{c}^\text{(a)}<\sigma_\text{c}^\text{(b)}$.  Naturally, however, for large $p$ as more and more links are rewired the local organisation is destroyed and the critical points of the two small world models converge.

To further strengthen the main assertion of the paper that feed forward loops facilitate synchronization, we also investigated a probabilistic variant of the small world model, for which the fraction of feed forward loops and cycles can be systematically tuned. For this, we consider the ring graph with $r=2$ and $l=3$ and set $a_{ji}=\delta_{j,i-1}+\delta_{j,i-2}$ with probability $q_\text{ff}$ and $a_{ji}=\delta_{j,i-1}+\delta_{j,i+1}$ otherwise. As implied by the notation the parameter $q_\text{ff}$ gives the ratio of feed forward loops and cycles.
\begin{figure}[tbp]
 \includegraphics[width=.45\textwidth]{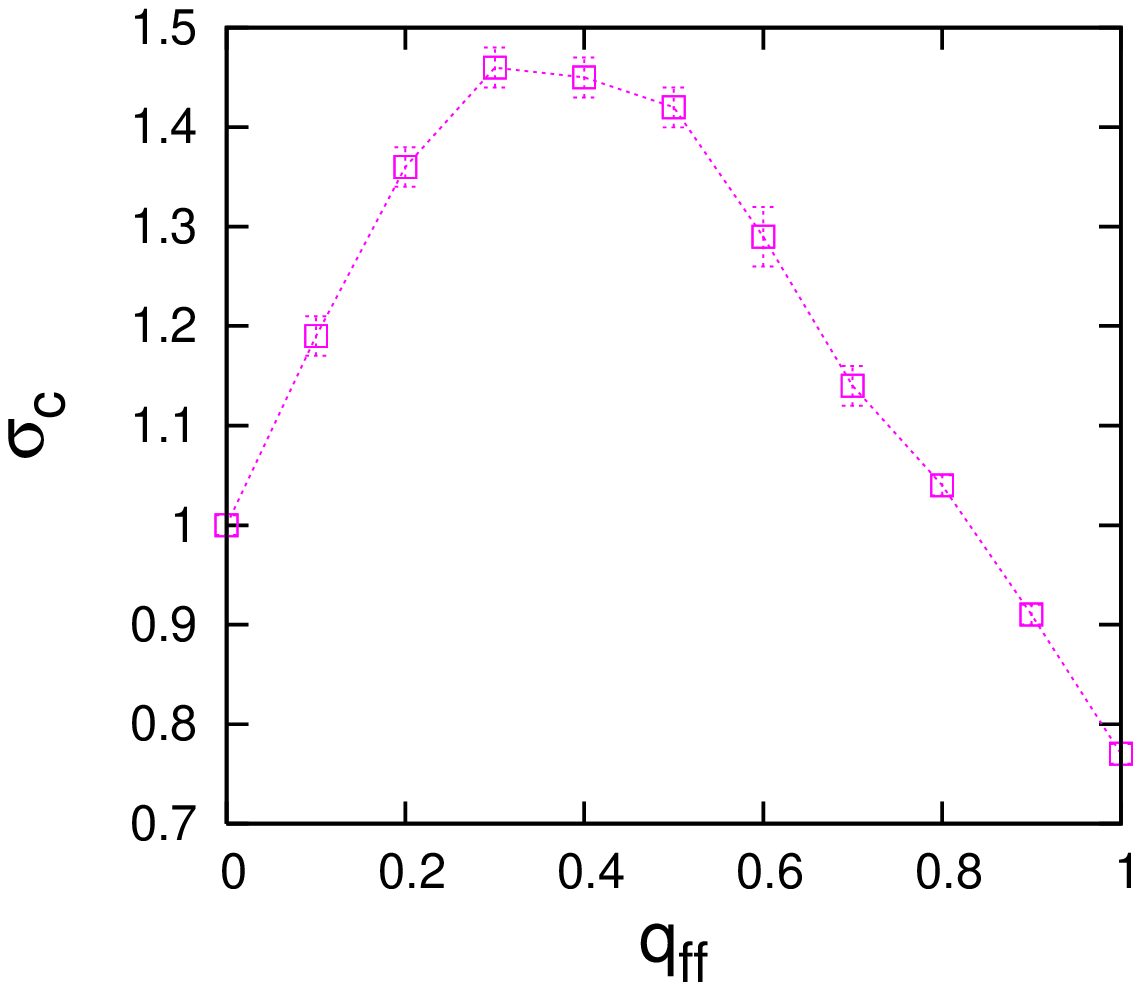} 
 \caption{Dependence of the critical point of the synchronization transition $\sigma_\text{c}$ on the ratio of feed forward loops to cycles $q_\text{ff}$ for small worlds with shortcut density $p=.2$. Results from FSS analysis, lines are guides for the eye.}
\label{fig.4a}
\end{figure}
One should, however, note that in this model the in-degree sequence of the graphs is no longer homogeneous. Instead, the variance of the in-degree sequence is $\sigma^2_{k_\text{in}}=2q_{\text{ff}}(1-q_{\text{ff}})$, which has a maximum for equal fractions of feed forward loops and cycles, i.e. for $q_\text{ff}=1/2$. Hence, as the in-degree heterogeneity strongly influences a network's synchronization properties, one has two competing tendencies when $q_\text{ff}$ is tuned. First, increasing $q_\text{ff}$ from zero onwards the in-degree variance of the graphs grows strongly. Second, increasing $q_\text{ff}$ also increases the fraction of feed forward cycles.

Both competing tendencies can be seen from the data in Fig. \ref{fig.4a}, which gives the dependence of the critical coupling $\sigma_\text{c}$ on $q_\text{ff}$. By the above discussed increase in in-degree variance, initially, replacing cycles by feed forward loops deteriorates the networks' synchronizabilities. The negative impact of the increased heterogeneity competes with the positve impact of the enhanced feed forward loop density. The data in Fig. \ref{fig.4a} verify that $\sigma_\text{c}$ reaches a maximum {\it before} the in-degree variance reaches its maximum.

Another point to note is that the in-degree variance is the same for graphs built with a given density $q=q_\text{ff}$ (for feed forward loops) or $q=1-q_\text{ff}$ (for cycles) of one of the motifs. Thus, comparing the synchronization threshold for graphs built with $q_\text{ff}=q$ and $q_\text{ff}=1-q$ excludes the effect of different in-degree heterogeneities. For this comparison Fig. \ref{fig.4a} clearly demonstrates that larger densities of feed forward loops always entail a lower synchronization threshold $\sigma_\text{c}$.

\begin{figure}[tbp]
 \includegraphics[width=.45\textwidth]{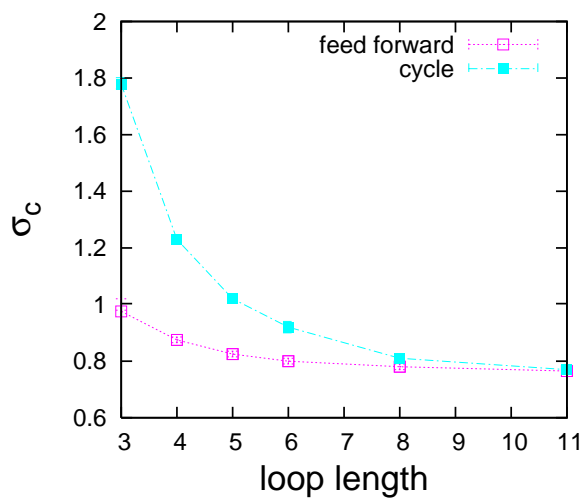} 
 \caption{Dependence of the critical point $\sigma_\text{c}$ on the length of loops for feed forward loops and cycles for small worlds with shortcut density $p=.05$. Results obtained from a FSS analysis. Error bars are below the size of the symbols.}
\label{fig.5}
\end{figure}

A natural extension of the conventional small world model with immediate nearest neighbour connections is to construct the ring graphs from loops of arbitrary length $l$, cf. Eq.'s (\ref{L1}) and (\ref{L2}). In this way the influence of general feed forward loops and cycles on the onset of synchronization can be explored. In Fig. \ref{fig.5} simulation results for the dependence of the critical point of small worlds with shortcut density $p=0.05$ on the loop length are displayed. 
Generally, introducing longer loops reduces average distances and thus favours synchronization. This is confirmed by a gradual decline of the critical points with increasing loop length in Fig. \ref{fig.5}. In agreement with the previous results for triangles one again observes that the graphs constructed from feed forward loops are more synchronizable than the graphs built from cycles.

The data of Fig. \ref{fig.5} indicate that the critical points for both models converge when loops become very long. One may surmise that for large $l$ the impact of shorter loops introduced by the shortcut links becomes dominant, thus obliterating the difference between the models.

\section{Conclusions}
In this paper we have studied the synchronization transition in several models of directed small world graph topologies. We stress the point that the synchronizablity of a small world strongly depends on the particulars of the local organization. Generally, having the same degree sequences and even though being larger in average distances, small world graphs built purely from feed forward loops are found to exhibit a transition towards full synchronization at a lower critical coupling strength than small worlds built purely from cycles. This point is corroborated by the study of the synchronization threshold of small worlds constructed with a tuneable ratio of feed forward loops and cycles. Other parameters being the same, small worlds with larger feed forward loop densities always exhibit a lower threshold to synchronization.

Similar to the spirit of Ref. \cite{Moreno} we also characterised the synchronizability of the standalone motifs. The finding that the more synchronizable motif gives rise to inferior synchronization properties globally cautions to conclude about global synchronization from local subgraph properties. It thus appears an appealing agenda for further research to study the influence of directed motifs on a network's synchronization properties.

\acknowledgments
I thank F. Boschetti for helpful discussions and comments about the manuscript and two anonymous referees for helpful comments.


\begin{thebibliography}{0}
\bibitem{Remark0}
 More generally, to systems with coupling matrices that can be mapped to a symmetric matrix with identical spectrum, see e.g. \cite{Hwang,Motter,Motter1,Chavez}.
\bibitem{Remark}	
        This rewiring procedure makes sure that the in-degree sequence of the graph is unaltered. We also experimented with other rewiring procedures and found no qualitative influence on the results presented below.
\end{thebibliography}
\end{document}